\documentstyle[preprint,aps]{revtex}

\begin{document}
\title{Quantum numbers of the V$^{3+}$ ion in V$_{2}$O$_{3}$ }
\author{R.J. Radwa\'{n}ski}
\address{Center for Solid State Physics, S$^{nt}$ Filip 5, 31-150 Krak\'{o}w,%
\\
Institute of Physics, Pedagogical University, 30-084 Krak\'{o}w, Poland.}
\author{Z. Ropka}
\address{Center for Solid State Physics, S$^{nt}$ Filip 5, 31-150 Krak\'{o}%
w, Poland.\\
email: sfradwan@cyf-kr.edu.pl}
\maketitle

\begin{abstract}
It is argued that the V$^{3+}$ ion in V$_2$O$_3$ should be considered as
described by quantum numbers S=1 and L=3. It results from Hund's rules and
means that the two d electrons form the highly-correlated electron system
with the importance of the intra-atomic spin-orbit coupling.



Receipt date by Phys.Rev.Lett. 18.09.2000
\end{abstract}

\date{(19.09.2000)}

V$_{2}$O$_{3}$ attracts the substantial scientific interest by more than 50
years. Despite of it there is still strong discussion about the description
of its properties and its electronic structure. There is a long-standing
controversy between an S=1 model without an orbital degeneracy [1] and the
S=1/2 orbitally degenerate model of Castellani et al. [2].

Recently a new ''orbitally degenerate spin-1 model for the insulating V$_2$O$%
_3$'' has been proposed in Ref. 3.

In all of these considerations it is agreed that V$_2$O$_3$ is
antifferomagnet with the Neel temperature of 160 K. It has the corundum Al$%
_2 $O$_3$ structure. It is agreed that V ions in the corundum structure sit
in an O octahedron with a small trigonal distortion. It is also agreed that
the V$^{3+}$ ion has two d electrons [1-3]. The base for all theories is the
description of the V$^{3+}$ ion and its electronic structure.

The aim of this Letter is to point out that the V$^{3+}$ ion in V$_2$O$_3$
should be described by the quantum numbers S=1 and L=3. It follows from the
Hund's rules (two i.e. the maximal S and maximal L for two d electrons) [4].

We understand that Hund's rules are phenomelogical rules but they are
well-established in physics. They reflect intra-atomic interactions. Thus,
the demand for keeping the Hund rules is equivalent to an idea that atoms
preserve their internal atomic structure being the part of a solid. The V$%
^{3+}$ ion has the 3d$^{2}$ configuration. It means that these two electrons
form the highly-correlated electronic system described by the resultant
quantum numbers S=1 and L=3.

The electronic structure in the octahedral crystal field and in the presence
of the intra-atomic spin-orbit coupling has been calculated by us in Ref. 5.

In conclusion, we claim that the V$^{3+}$ ion in V$_2$O$_3$ should be
considered as described by the quantum numbers S=1 nad L=3. We are convinced
that taking into account the orbital moment and the intra-atomic spin-orbit
coupling is indispensable for the description of electronic and magnetic
properties of V$_2$O$_3$.

Email for correspondence: sfradwan@cyf-kr.edu.pl

\end{document}